\documentclass[aps,preprint,showpacs]{revtex4}
\usepackage{psfig}
\begin{document}        %
\draft
\title{Effect of Corrugations to steady thermal explosion \\ with  power-law
thermal absorptivity in microannuli}  %
\author{Zotin K.-H. Chu$^{1}$ and Chen Qin$^2$}
\affiliation{$^1$ 3/F, 4, Alley 2, Road Xiushan, Leshanxinchun,
Xujiahui 200030, China \\ 
$^2$ Department of Physics, Xinjiang Normal University, Urumqi
830054,
China  
}
\begin{abstract}
We obtain the approximate solutions for the steady temperature
profiles of materials with a temperature-dependent thermal
absorptivity inside a microannulus with wavy-rough surfaces
considering a quasilinear partial differential equation by the
boundary perturbation approach.  Our numerical results show that the
critical Frank-Kamanestkii  parameter  depends on the
small-amplitude  wavy-roughness for specific power-law of thermal
absorptivity.

%
\end{abstract}
\maketitle
%
\bibliographystyle{plain}
\section{Introduction}
Orbiting spacecraft are subjected to forces such as drag and solar
pressure that have to be compensated for in order to maintain the
desired orbit, and they may also have to be transferred from the
initial orbit where they are placed by the launcher to their final
orbit. The orientations and orbits of spacecraft are controlled by
low-thrust propulsion systems. The propulsion control systems can be
broadly classified as suitable for coarse and fine control.
Hitherto, the major effort has been expended on the development of
coarse control systems. In the 1960s, simple cold gas propulsion
systems were employed and over the years more efficient systems were
developed such as monopropellant hydrazine fuel, storable
bipropellant mixtures providing more efficient combustion and, quite
recently, new type of propulsion systems to provide higher
propulsive impulse and higher performance to cope with increased
spacecraft masses and longer satellite lifetimes [1-2]. The crucial
issue which is related to the  important components of many space
systems is the thruster system that shall generate thrust in a
well-defined direction. One means for thrust generation is
solid-state micro-thrusters which requires the safety storage of
fuels [3].\newline
Meanwhile, thermal instability such as the catastrophic phenomenon
of thermal runaway in which a slight change of external heating
(e.g., microwave)  power causes the temperature to increase rapidly
to the melting point of the material (e.g., ceramic) have been
studied for simple geometry cases [4-10]. Up to now, the basic
understanding of the (microwave) heating process still remains
somewhat empirical and speculative due to its highly nonlinear
character. The mathematical description of this process is
accordingly fraught with nonlinearities (not to mention the external
heating) : The heat equation is highly temperature dependent at the
temperatures required for sintering. Moreover, the thermal boundary
conditions must take into account both convective and radiation heat
loss. The result is a highly nonlinear initial-boundary value
problem. Even for simple geometry problems, because of the highly
nonlinearity characteristic, numerical approaches were often adopted
to solve the corresponding equations and boundary conditions
[4,9-10]. \newline Relevant approaches have also a wide field of
applications in combustion theory or  for safety of storage of
materials capable of exothermic chemical reaction [11]. In this
short paper, we shall consider the above mentioned problems with
simple geometry (say, an annulus) in microdomain. However, real
surfaces are rough at the micro- or even at the meso-scale [12-14].
We presume the roughness to be wavy-like in transverse direction
along the outer and inner surfaces of the annulus which thus make
the corresponding boundary conditions highly nonlinear. With this,
we adopt the boundary perturbation approach [12-14] to handle the
mathematically complicated boundary value problems. Our numerical
results show that for certain power-law of  thermal absorptivity the
small amplitude wavy-roughness could influence the steady thermal
behavior inside a microannulus significantly.
\section{Mathematical  formulations}
Considering the forced heat equation
\begin{displaymath}
 \rho c_p(T) \frac{\partial T}{\partial t}=\nabla \cdot
 \biggl(\nu(T) \nabla T\biggr)
 +\gamma(T) \phi^2
\end{displaymath}
where $T$ is the temperature,  $\nu$ is the thermal conductivity,
$\gamma$ is the thermal absorptivity, $\rho$ is the density that
is usually assumed constant,  $c_p$ is the specific heat and
$\phi$ is the heating source. Here, the thermal absorptivity
depends on the square of the external source (amplitude). Assuming
constant specific heat and that the heating source  (amplitude) is
temperature independent, the appropriate form of above equation
becomes
\begin{equation}
  \frac{\partial T}{\partial t}=\nabla \cdot
 \biggl(\kappa(T) \nabla T\biggr)
 +\gamma(T) \phi^2,
\end{equation}
where $\phi$ might depend on the spatial coordinates.
\subsection{Smooth Cylinder Geometry}
Without loss of the generality, we consider a class of solutions
for the nonlinear reaction-diffusion equations
\begin{equation}
  \frac{\partial T}{\partial t}=\frac{\partial}{x\partial x }
 \biggl(x T^m \frac{\partial T}{\partial x} \biggr)
 +\delta K(x) T^n
\end{equation}
with the geometry of an infinite cylinder ($x\equiv r$ with $r$
being the radius) and $m=-1$. Here, $\delta$ is the
Frank-Kamanestkii parameter [15] which incorporates the chemical
properties of the combustible material, the temperature of assembly
as well as its geometrical dimensions. $K(x)=x^{\mu}$ and $\mu <0$,
$\mu$ is  a constant. The critical value $\delta_{cr}$ defines
criticality and provides, for example, a criterion for safety of
storage of materials capable of exothermic chemical reaction
[11,15]. Thus, $\delta_{cr}$ is important in characterizing the
thermal stability properties of the material under consideration.
\newline For the smooth geometry case, we consider the steady state thermal explosions
between two infinite cylindrical material undergoing exothermic
reaction by using the change of variable given by $x=\exp(y)$,
$T^n=\exp(U)$ and $\Theta=U+(2+\mu)y$ and we can obtain that
closed-form solutions which are available for all real values of
$\mu$ [15]. In fact, equation (2) becomes
\begin{equation}
 \frac{d^2 \Theta}{d y^2}+\delta n e^{\Theta}=0
\end{equation}
and the solutions are
\begin{equation}
 e^{\Theta}=\frac{C_0}{\cosh^2
 \biggl(\sqrt{\frac{n\delta C_0}{2}}y-D_0 \biggr)} \hspace*{12mm}
 \mbox{or} \hspace*{12mm} T^n=\frac{C_0 x^{-2-\mu}}{\cosh^2
 \biggl(\sqrt{\frac{n\delta C_0}{2}}\ln x -D_0 \biggr)}
\end{equation}
where $C_0$, $D_0$ are integration constants.\newline Note that
for the pure cylindrical case ($x\equiv r=0$), once $\mu \ge -1$,
we must make sure the solution to be well defined considering the
limit of $x\rightarrow 0$ :
\begin{equation}
 T^n=\frac{(\mu+2)^2}{2n \delta x^{2+\mu}\cosh^2
 \biggl({\frac{2+\mu}{2}} \ln x-D_0 \biggr)} 
\end{equation}
with
\begin{displaymath}
 C_0=\frac{(2+\mu)^2}{2 n\delta}.
\end{displaymath}
The boundary conditions for a smooth annulus could be
\begin{equation}
 T^n (1)=1 \hspace*{12mm} \mbox{and} \hspace*{12mm} T^n(s)=1,
\end{equation}
where $s=r_1/r_2$ (cf. Fig. 1) and $x=r/r_2$. These could help us
to fix $C_0$ and $D_0$. Under this situation, we have
\begin{equation}
 \sqrt{\frac{n \delta}{2}} \ln s \, \cosh
 (-D_0)=D_0+\cosh^{-1} [s^{\frac{-\mu-2}{2}} \cosh(-D_0)],
\end{equation}
where $\mu < -1$.
\subsection{Wavy-Rough Annulus Geometry}
We start to consider a steady state heat flow  in a wavy-rough
microannulus of $r_2$ (in mean-averaged outer radius) with the outer
wall being a fixed wavy-rough surface : $r=r_2+\epsilon \sin(k
\theta)$ and $r_1$ (in mean-averaged inner radius) with the inner
wall being a fixed wavy-rough surface : $r=r_1+\epsilon \sin(k
\theta+\beta)$, where $\epsilon$ is the amplitude of the (wavy)
roughness, and the wave number : $k=2\pi /L $ ($L$ is the wave
length), $\beta$ is the phase shift. The schematic is illustrated in
Fig. 1. \newline With the boundary perturbation approach  from Refs.
[12-14], we shall derive the temperature field along the wavy-rough
microannuli considering dimensionless values. We firstly select
$L_a\equiv r_2$ to be the characteristic length scale and set
$$r'=r/L_a, \hspace*{5mm}
R_o=r_2/L_a, \hspace*{5mm} R_i=r_1/L_a, \hspace*{5mm}
\epsilon'=\epsilon/L_a. $$ After this, for simplicity, we drop all
the primes. It means, now, $r$, $R_o$, $R_i$, and $\epsilon$
become dimensionless. The walls are prescribed as $r=R_o+\epsilon
\sin(k\theta)$, $r=R_i+\epsilon \sin(k \theta+\beta)$ and the
presumed steady state heat flow is along the $z$-direction
(microannulus-axis direction). 
\newline
Let $T$ be expanded in $\epsilon$ :
\begin{equation}
 T= T_0 +\epsilon T_1 + \epsilon^2 T_2 + \cdots,
\end{equation}
and on the boundary, we expand $T(r_0+\epsilon dr,
\theta(=\theta_0))$ into
\begin{displaymath}
T(r,\theta) |_{(r_0+\epsilon dr ,\theta_0)}
=T(r_0,\theta)+\epsilon [dr \,T_r (r_0,\theta)]+ \epsilon^2
[\frac{dr^2}{2} T_{rr}(r_0,\theta)]+\cdots
\end{displaymath}
where the subscript means the partial differentiation (say, $T_r
\equiv
\partial T/\partial r$). \newline
The corresponding equation, for simplicity (as the wavy roughness
being imposed) as we select firstly $n=m+1$, becomes
\begin{equation}
 \frac{\partial}{r\partial r }
 \biggl(r T^m \frac{\partial T}{\partial r} \biggr)
 +\delta r^{\mu} T^{m+1} =0
 \end{equation}
or
\begin{equation}
  \frac{\partial}{r\partial r }
 \biggl(r  \frac{\partial \Theta}{\partial r} \biggr)
 +(m+1)\delta r^{\mu} \Theta =0
\end{equation}
with $\Theta=T^{m+1}$. \newline Now, we can set
$\eta=r^{(2+\mu)/2}$ and then obtain
\begin{equation}
 \frac{\partial^2 \Theta}{\partial \eta^2}+\frac{\partial
 \Theta}{\eta \partial \eta}+\frac{4(m+1)\delta}{(2+\mu)^2}
 \Theta =0, \hspace*{12mm} \mu \not=-2.
\end{equation}
Above equation has the zeroth order solution in terms of Bessel
functions
\begin{equation}
 \Theta_0 = A_0 J_0 (K \eta)+ B_0 Y_0 (K \eta),
\end{equation}
once we consider $\Theta=\Theta_0+\epsilon \Theta_1 +\cdots$ [16].
Here, $A_0$, $B_0$ are integration constants. $K=2\sigma/(2+\mu)$
and $\sigma=[(m+1)\delta]^{1/2}$. Thus, we obtain, for $\mu \not =
-2$,
\begin{equation}
 \Theta_0 = T_0^n =T_0^{m+1}=A_0 J_0
 \biggl(\frac{2[(m+1)\delta]^{1/2}}{2+\mu} r^{(2+\mu)/2}\biggr)+
 B_0 Y_0
 \biggl(\frac{2[(m+1)\delta]^{1/2}}{2+\mu} r^{(2+\mu)/2}\biggr)
\end{equation}
and for $\mu=-2$,
\begin{equation}
  \Theta_0 = T_0^n =T_0^{m+1}=A^* \sin\biggl([(m+1)\delta]^{1/2} \ln
  r\biggr)+B^* \cos\biggl([(m+1)\delta]^{1/2} \ln r\biggr).
\end{equation}
Here, $A_0$,$B_0$ or $A^*$,$B^*$ could be uniquely determined
after imposing the boundary conditions.
 This leads to $T_0=\Theta_0^{1/(m+1)}$.
\newline
As for the first order correction or solution, with the
wavy-roughness effect, we can only have (considering
$T^n\equiv(T_0+\epsilon T_1+\cdots)^n$$=T_0^n+\epsilon n T_1
T_0^{n-1}+\cdots$ with $T_0$ known), for $m=0$,
\begin{equation}
 \frac{\partial^2 T_1}{\partial r^2}+\frac{\partial
 T_1}{r \partial r}+\frac{1}{r^2}\frac{\partial^2 T_1}{\partial \theta^2}
 +\delta r^{\mu} T_1 =0, 
\end{equation}
due to the complicated orientational characteristics
($\theta$-dependent).
 Using the boundary conditions from the equation (6), we have
\begin{equation}
 T_0|_{r=s}=1, \hspace*{12mm} T_0|_{r=1}=1,
\end{equation}
\begin{equation}
 \frac{\partial T_0}{\partial r}|_{r=1} \sin (k\theta)+T_1|_{r=1}=0,
\end{equation}
and
\begin{equation}
 \frac{\partial T_0}{\partial r}|_{r=s} \sin (k\theta+\beta)+T_1|_{r=s}=0,
\end{equation}
Equation (15) can be transformed to
\begin{equation}
  \frac{\partial^2 R}{\partial \xi^2}+\frac{\partial
 R}{\xi \partial \xi}+(\frac{2}{2+\mu})^2(\delta r^{\mu+2}-K^2) R
 =0, \hspace*{12mm}
T_1 = i e^{iK\theta} R(\xi),
\end{equation}
and solved after setting
\begin{equation}
 \xi^2=r^{\mu+2}, \hspace*{12mm} \nu=\frac{2}{2+\mu}\delta^{1/2},
 \hspace*{12mm} \bar{K}=\frac{2}{2+\mu}K, \hspace*{12mm}
 \mu\not=-2,
\end{equation}
where
\begin{equation}
 R(\xi)=A_1 J_{\bar{K}}(\nu \xi)+B_1 J_{-\bar{K}} (\nu \xi).
\end{equation}
We thus have
\begin{equation}
 T_1 = i e^{iK\theta} \biggl[A_1 J_{\frac{2}{2+\mu}K}
 \biggl(\frac{2}{2+\mu} \delta^{1/2} r^{(2+\mu)/2}\biggr)+B_1
 J_{-\frac{2}{2+\mu}K}
 \biggl(\frac{2}{2+\mu} \delta^{1/2} r^{(2+\mu)/2}\biggr)\biggr],
\end{equation}
with $\mu\not=-2$ or
\begin{equation}
 T_1 = i e^{iK\theta} \biggl[C_1 \cos\biggl((\delta-K^2)^{1/2}\ln
 r\biggr)+D_1 \sin\biggl((\delta-K^2)^{1/2}\ln
 r\biggr)\biggr], \hspace*{12mm} \mu=-2.
\end{equation}
Here, $A_1$,$B_1$,$C_1$,$D_1$ could be fixed by boundary
conditions or equations (16-18). In fact, from the boundary
condition (17) or the wavy roughness along the annulus, we can
identify $K=k$. \newline
With the first order perturbed solution due to wavy roughness :
$T_1$, we can thus obtain the approximate solution for especially
$m=0$
\begin{equation}
 T=T_0 + \epsilon T_1 +\cdots,
\end{equation}
where
\begin{equation}
 T_0=A_0 J_0
 \biggl(\frac{2\delta^{1/2}}{2+\mu} r^{(2+\mu)/2}\biggr)+
 B_0 Y_0
 \biggl(\frac{2\delta^{1/2}}{2+\mu}
 r^{(2+\mu)/2}\biggr), \hspace*{12mm} \mu \not =-2
\end{equation}
or
\begin{equation}
 T_0=A^* \sin\biggl(\delta^{1/2} \ln
  r\biggr)+B^* \cos\biggl(\delta^{1/2} \ln r\biggr), \hspace*{12mm} \mu
  =-2,
\end{equation}
with $T_1$ given by equations (22) and (23). To be specific, we
have ($s \equiv R_i$)
\begin{displaymath}
 A_0 =\frac{Y_0 (R_s s^{(2+\mu)/2})- Y_0 (R_s)}{J_0 (R_s)\,Y_0 (R_s s^{(2+\mu)/2})-Y_0 (R_s) J_0 (R_s
 s^{(2+\mu)/2})},
\end{displaymath}
\begin{equation}
 B_0 =\frac{J_0 (R_s)- J_0 (R_s s^{(2+\mu)/2})}{J_0 (R_s)\,Y_0 (R_s s^{(2+\mu)/2})-Y_0 (R_s) J_0 (R_s
 s^{(2+\mu)/2})},
\end{equation}
where $R_s = {2}\sqrt{\delta}/({2+\mu})$, and
\begin{equation}
 A^* =\frac{1- \cos(\sqrt{\delta} \ln s)}{\sin(\sqrt{\delta} \ln s)}, \hspace*{12mm}  B^* =1.
\end{equation}
Meanwhile, with equations (17) and (18), we can obtain
\begin{equation}
 A_1=\biggl[F_1 J_{\frac{-2K}{2+\mu}} \biggl(\frac{2\sqrt{\delta}}{2+\mu}
s^{(2+\mu)/2} \biggr) - G_1 J_{\frac{-2 K}{2+\mu}}
\biggl(\frac{2\sqrt{\delta}}{2+\mu}\biggr)\biggr]/det,
\end{equation}
\begin{equation}
 B_1=\biggl[G_1 J_{\frac{2}{2+\mu}K} \biggl(\frac{2\sqrt{\delta}}{2+\mu}
 \biggr) - F_1 J_{\frac{2}{2+\mu}K}
\biggl(\frac{2\sqrt{\delta}}{2+\mu}s^{(2+\mu)/2}\biggr)\biggr]/det,
\end{equation}
where
 $$ F_1 =-\sqrt{\delta} \biggl[A_0 J_1
(\frac{2\sqrt{\delta}}{2+\mu})+B_0 Y_1
(\frac{2\sqrt{\delta}}{2+\mu})\biggr],$$
 $$ G_1 =-e^{i\beta}\sqrt{\delta} s^{\mu/2}\biggl[A_0 J_1
(\frac{2\sqrt{\delta} s^{(2+\mu)/2}}{2+\mu})+B_0 Y_1
(\frac{2\sqrt{\delta} s^{(2+\mu)/2}}{2+\mu})\biggr],$$
$$ det=J_{\frac{2}{2+\mu}K} \biggl(\frac{2\sqrt{\delta}}{2+\mu}\biggr)
J_{\frac{-2K}{2+\mu}} \biggl(\frac{2\sqrt{\delta}}{2+\mu}
s^{(2+\mu)/2} \biggr)-J_{\frac{-2 K}{2+\mu}}
\biggl(\frac{2\sqrt{\delta}}{2+\mu}\biggr) J_{\frac{2}{2+\mu}K}
\biggl(\frac{2\sqrt{\delta}}{2+\mu} s^{(2+\mu)/2} \biggr).$$
\section{Numerical Results and Discussions}
We plot $T_0(r)$ for different $\mu$ ($=-4,-1,1$) with fixed $\delta
(=2)$ in Fig. 1. The boundary conditions : $T_0
(r=s=0.4)=T_0(r=1)=1$ are correctly matched at both walls.
Considering the contributions from the first-order perturbation $T_1
(r)$, we plot the approximate temperature profiles : $T_0
(r)+\epsilon T_1 (r)$ with the same $\mu(=-4,-1,1)$ in Fig. 2. Here,
$\epsilon=0.1$, $\beta=\pi/4$ and $k=8$. We can observe that, for
$\mu=-4$, the effect of first-order perturbation is quite large
compared to other cases of  $\mu$.
\newline
To consider the effects of roughness, we could observe that, as
evidenced from the first-order correction or approximation : $T_1$,
the Frank-Kamanestkii parameter ($\delta$) depends on the wave
number of wavy-roughness (cf. Eq. (23)). Thus,  the critical
Frank-Kamanestkii $\delta_{cr}$ depends on the square of the wave
number $k$ (or $k_{cr}$ for certain critical ones). However, above
results are only for cases of $\mu=-2$ (or $K(x)=x^{-2}$ which is
defined in Eq. (2))!

\section{Conclusion}
We adopt the boundary perturbation approach to obtain the
approximate solutions for the steady temperature profiles  of
materials with a temperature-dependent thermal absorptivity inside a
microannulus with wavy-rough surfaces considering a quasilinear
partial differential equation.  Our results show that the critical
Frank-Kamanestkii  parameter strongly depends on the wavy-roughness
for specific power-law of thermal absorptivity. {\it
Acknowledgement.} 
The first author stayed at the Chern Shiing-Shen Institute of
Mathematics, Nankai University for one month. Thus the first author
should thank their hospitality for their Visiting-Scholar Program.
%


\newpage

\psfig{file=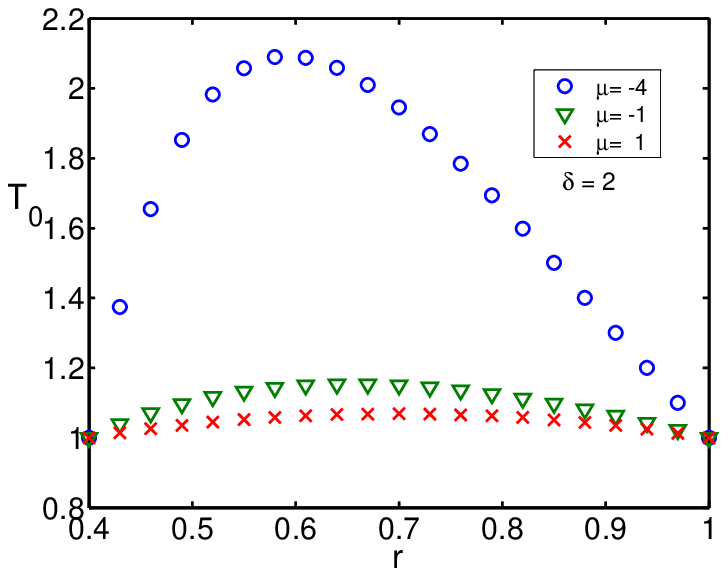,bbllx=-1.0cm,bblly=18.5cm,bburx=10cm,bbury=26.5cm,rheight=8cm,rwidth=10cm,clip=}

\begin{figure}[h]
\hspace*{10mm} Fig. 1. Approximate zeroth order temperature profiles
: $T_0 (r)$ for different $\mu$ ($=-4,-1,1$) with $\delta=2$. The
boundary conditions are satisfied at both sides ($T_0
(s=0.4)=T_0(1)=1$).
\end{figure}

\newpage


\psfig{file=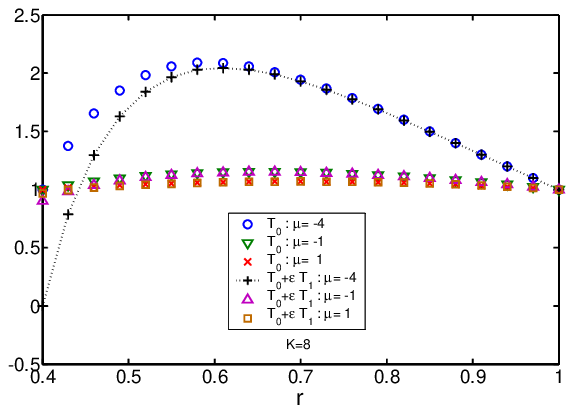,bbllx=-1.0cm,bblly=22cm,bburx=10cm,bbury=30cm,rheight=8cm,rwidth=10cm,clip=}

\begin{figure}[h]
\hspace*{10mm} Fig. 2. Approximate $T_0 (r)+\epsilon T_1 (r)$
temperature profiles for different $\mu$ ($=-4,-1,1$) with
$\delta=2$. Here, the wave number of roughness $k$ is equal to $8$,
$\epsilon=0.1$, $\beta=\pi/4$ and the inner boundary is $s=0.4$. We
can observe that the small-amplitude wavy-roughness changes the
near-inner-wall temperature distribution significantly once
$\mu=-4$.
\end{figure}
\end{document}